# Improving Productivity through Corporate Hackathons: A Multiple Case Study of Two Large-scale Agile Organizations

Nils Brede Moe
SINTEF
nils.b.moe@sintef.no

Rasmus Ulfsnes
SINTEF
Rasmus.Ulfsnes@sintef.no

Viktoria Stray
University of Oslo, SINTEF
stray@ifi.uio.no

Darja Smite
SINTEF, BTH
darja.smite@bth.se

**Abstract**

*Software development companies organize hackathons to encourage innovation. Despite many benefits of hackathons, in large-scale agile organizations where many teams work together, stopping the ongoing work results in a significant decrease in the immediate output. Motivated by the need to understand whether and how to run hackathons, we investigated how the practice affects productivity on the individual and organizational levels. By mapping the benefits and challenges to an established productivity framework, we found that hackathons improve developers' satisfaction and well-being, strengthen the company culture, improve performance (as many ideas are tested), increase activity (as the ideas are developed quickly), and improve communication and collaboration (because the social network is strengthened). Addressing managerial concerns, we found that hackathons also increase efficiency and flow because people learn to complete work and make progress quickly, and they build new competence. Finally, with respect to virtual hackathons we found that developers work more in isolation because tasks are split between team members resulting in less collaboration. This means that some important, expected hackathon values in virtual contexts require extra effort and cannot be taken for granted.*

## 1. Introduction

The phenomenon of corporate hackathons has emerged as a popular approach for encouraging innovation within established software development companies [1], inspired by the collegiate and civic events [1, 2]. The word originates from the combination of the words "hack" and "marathon" [3], where developers use a limited and defined amount of time to create a minimum viable product (or at least something to show at a demo) [4]. While the participants and audience are often satisfied with the outcome of the hackathon, results are rarely exploited commercially [2, 5]. Consequently, the most significant value from hackathons is the opportunity for people to meet and collaborate to create new networks within the organization and not to test new ideas quickly [3].

During a hackathon, employees involved stop working on their regular tasks for two or more days. Some companies conduct several hackathons a year. In large-scale agile organizations where many teams work together to deliver software and thus are dependent on each other [6, 7], stopping the whole or parts of the organization will impair the immediate plans of delivering code to production. Therefore, it is essential for participants and organizers to be aware of the benefits and costs of hackathons on the organizational level, even though pursuing individual goals is considered an entirely legitimate use of hackathon time [2]. While the number of hours spent on hackathons is easily calculated, the invisible gains and potential losses are not yet understood well.

Positive effects of hackathons are empowerment, having fun with colleagues, expanding competencies, networking among developers, and identifying new ideas for the company [8-11]. Previous research shows that freedom and fun make software developers productive and engaged [12]. Furthermore, testing new ideas quickly might speed up the innovation process. Learning through experimentation is one fundamental prerequisite for agile and innovative software development [13]. One could argue that even if hackathons do not result in new businesses, it is still a key practice that every software company needs to use. While the benefits are numerous, calculating the actual gains for an organization requires a better understanding of how these benefits contribute to productivity.

Productivity in software engineering has been discussed for decades. Yet, the questions of how to measure or even define developer productivity remain elusive [14]. Forsgren et al. [14] argue that productivity cannot be reduced to a single metric and need to be considered on five dimensions: satisfaction and well-being; performance; activity; communication and collaboration; and efficiency and flow. From the existing research on hackathons, it is likely that the practice positively influences several of these five dimensions. However, as most research on hackathons focuses on collegiate or civic events [1, 2], there is a need for research on corporate hackathons when understanding the impact on productivity.

Motivated by the need to understand how hackathons affect productivity in large organizations, and to better guide organizations when deciding on the use of hackathons, we ask (*RQ1): How do hackathons affect productivity in large-scale agile?*

The well-being of developers has never been more critical than in the Covid-19 pandemic, which has forced most software developers worldwide to work from home. Therefore, one could argue that the role of hackathons is even more critical, as the event allows colleagues to have fun [8, 15], and developers are more satisfied when they have autonomy over their work [16]. Yet, little is known about virtually held hackathons. Therefore, we ask (*RQ2): How do developers experience hackathons held virtually when working from home?*

To answer these questions, we performed a multiple case study on hackathons in two large-scale agile companies that performed hackathons before and during the Covid-19 pandemic. Furthermore, we relied on the SPACE framework [14] to capture the different dimensions of productivity. Our study extends the single case study reported in [15] by adding another case and a research question on productivity.

The remainder of this paper is organized as follows: We present background on hackathons and developer productivity in Section 2. In Section 3, we describe the two case companies and the research method that we used. In Section 4, we present the results. In Section 5, we discuss the findings, suggest implications for practice, and discuss the study's limitations. Section 6 concludes the paper and proposes future work.

## 2. Background

### 2.1. Hackathons

Large Agile organizations have utilized hackathons increasingly as a mechanism for employee empowerment and innovation in the last 20 years [4, 5, 8]. Hackathons appeared first with OpenBSD and SUN Microsystems in 1999 [3]. There are multiple alternative names for hackathons, such as hackfest, jam, codefest, bug bash [5], as well as other more obscure names such as Delivery Day [8] and FedEx Day [17]. Hackathons can be open events organized by universities, cities, or municipalities, or they can be internal corporate events.

Valenca et al. [4] found that corporate hackathons follow a ten-step process (see Figure 1). The pre-hackathon phase involves a decision on aspects, such as (i) whether the event will be internal, external, or hybrid; (ii) whether the event will last for a few days or a whole week; or (iii) the venue. The hackathon is initiated with an opening ceremony. This activity involves the themes or challenges to be addressed, the rules, instructions, expectations, and the hackathon process. When the hackathon prototypes are developed, the event finishes with the closing ceremony. The closing ceremony consists of the teams' demonstration of their results, which a group of evaluators might then judge. This step may vary in terms of team selection, jury composition and voting process. In the post-hackathon phase, organizers may assess participants' satisfaction with the hackathon.

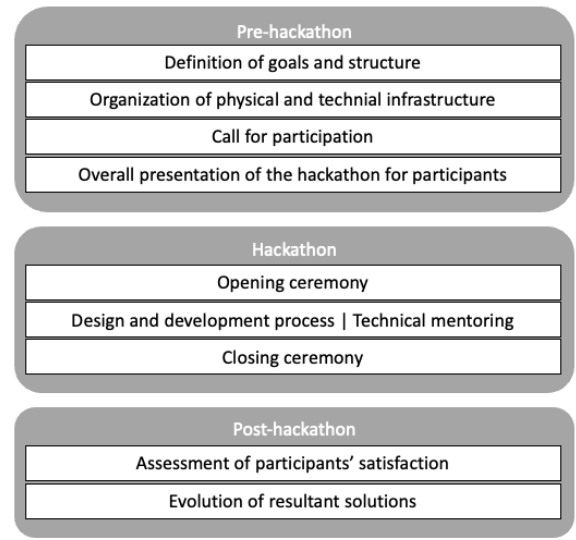

**Figure 1 Phases and activities in hackathons [4]**

The characteristics of hackathons provide different benefits and challenges. Examples of challenges are prototypes that do not receive sufficient follow-up work [1, 2, 9] and the stress of putting regular work aside [8]. Contradictions and tradeoffs are something apparent in the hackathon phenomenons. For example, the composition of a cross-disciplinary hackathon team benefits learning and networking and is a potential hinderance for effective work during the hackathon [2].

Falk Olesen and Halskov [1] emphasize that hackathon organizers need to tune its characteristics to achieve the wanted benefits for the target group. Nolte et al. [2] argue that participants and organizers should be aware of the tradeoff between developing a product that will continue after the hackathon, and achieving individual goals or working on radically new ideas that do not fit any existing product lines.

While recommended processes for hackathons exist, the events vary in how they are organized and executed, how ideas are structured, whether or not it is a competition and where they occur (physical or virtual) [11]. Due to the Covid-19 pandemic, corporate hackathons have been increasingly organized as virtual events through MS Teams, Google Meet, Zoom, and Slack. Furthermore, as many companies have announced their *Work from anywhere* (WFX) strategy (for example, Twitter, Spotify, Facebook, Salesforce), such events will likely be organized virtually in the future, as well.

## 2.2. Hackathons as an empowerment strategy

Empowerment and autonomous teams are critical requisites for succeeding with large-scale agile. While strategies and practices are reported in companies such as Zappos [18] and Spotify [19], autonomy and team empowerment is challenging in large-scale agile [19, 20]. Hackathons help companies empower their employees by granting days to work on delivering a software product idea of their choice [4] together with freely selected team members. When team members are given control over their work, scheduling and implementing their tasks, the result is increased motivation, job satisfaction [16], innovation, and performance.

One potential challenge with organizing a hackathon once or twice a year is that it is seldom sufficient for developers to feel empowered continuously. Therefore, ideally hackathons need to co-exist with other practices following the same principles of empowerment [8]. If the hackathon is the only event during a year, it is likely to fail to have a lasting effect on employee empowerment. Some companies grant employees additional freedom by introducing innovation programs [21], where employees can suggest and work on their ideas, and 20% time programs as implemented by Google and Atlassian [17], where developers get 20% of their time to work on a project or initiative of their choosing. At Atlassian, 20% time and hackathons are seen as two complementary practices that bring about different types of innovations important to the company [17].

## 2.3. Developer productivity and hackathons

Translating the gains from organizing corporate hackathons into individual or organizational productivity is paramount for motivating companies to make the needed investment into practicing hackathons regularly. Yet, doing so is not a straightforward task because there is no single productivity metric for software development. The main reason is that it is almost impossible to measure and quantify all the facets of developer activities when developing software [14]. Consequently, at Google, they argue that there is a need to focus on a set of custom metrics targeted to a specific question [22]. In contrast, others suggest having a holistic approach and treating productivity as a complex and compound phenomenon [14]. In our research, we draw upon the SPACE framework, which associates productivity with five core dimensions across three levels (individual, team, system) [14]:

**Satisfaction and well-being:** "Satisfaction is how fulfilled developers feel through their work, team, tools, or culture; well-being is how healthy and happy they are, and how their work impacts it" [14]. Further productivity and satisfaction are correlated.

**Performance:** "Performance is the outcome of a system or process" [14]. As teams, not individuals write software, the performance of software developers is hard to quantify because it can be challenging to tie individual contributions directly to product outcomes. For this reason, productivity is measured as outcome, not output.

**Activity:** "Activity is a count of actions or outputs completed in the course of performing work" [14]. Developer activity can provide valuable insights about developer productivity, engineering systems, and team efficiency.

**Communication & collaboration:** "Communication and collaboration capture how people and teams communicate and work together" [14]. Software development is a collaborative, creative task that relies on effective communication, coordination, and collaboration within and between teams. Examples are networking (who is connected to whom) and onboarding.

**Efficiency and flow:** "Efficiency and flow capture the ability to complete work or make progress on it with minimal interruptions or delays, whether individually or through a system" [14]. Efficiency and flow can include how well activities within and across teams are orchestrated and whether continuous progress is made.

Forsgren et al. [14] suggest that all the five mentioned dimensions should be evaluated and balanced on the individual, group, and system levels.

While some of the benefits of hackathons [2, 8-11, 17, 23] are clearly related to the SPACE framework (such as having fun with colleagues, producing new ideas, and evaluating them quickly), others are not easy to relate to productivity (such as when individual participants want to learn one specific technology or skill, disregarding whether that is useful for the organization).

## 3. Research methodology and approach

We applied a multiple case study [24] to investigate how hackathon relates to productivity, and how hackathons are affected by work from home. The context of our research consists of two agile software companies that were a part of a research program on employee-driven innovation and team autonomy. The companies are identified here as "MobComp" and "FinCompTech" due to confidentiality.

**MobComp** is a leading app developer and content platform focusing on mobile phone personalization and entertainment. The MobComp mobile app currently has 500 million installs, 35 million monthly active users and 750.000 subscribing users. The organization has 50 employees with teams that span Europe and the US. Work methodology is Scrum of Scrum with a focus on team autonomy, with direction from Product Managers.

**FinCompTech** is a software company developing financial services in Norway. Twenty-four agile teams deliver services that include pension, savings, insurance and banking products, serving both the private and business market. The teams have considerable freedom in how they work. Most use a Kanban variant with Scrum elements.

We conducted a two-phase study to explore the different aspects of hackathons. In phase 1, we focused on exploring the benefits and challenges of hackathons in MobComp and how they are affected by working from home. In this phase, we first analyzed a demo from a recent hackathon in the company. Then, based on the demo and previous research [17] [8], we developed an interview guide covering the following areas: *Background, motivation, idea generation, cooperation, hackathon organization, expectations, benefits, challenges, virtual.*

Eight semi-structured interviews with people from four different teams in MobComp were conducted by the first and third authors. The interviewees represented various skills and roles within the company, *including* developers, lead architects, senior engineers, and an advertising operations manager. The interviews were conducted in Norwegian or English, recorded, transcribed and qualitatively coded using descriptive coding, followed by a holistic overview and a thematic analysis [25]. Finally, to validate the qualitative findings and elicit additional data, we surveyed 23 employees from MobComp to rate the benefits and challenges.

In phase 2, we used the emergent findings in Phase 1**,** where a potential link toward SPACE [14] was discovered. We reviewed documentation from FinCompTech about their hackathon practices and history. During the initial investigation of FinCompTech, we found that permanent employees had the opportunity to use 20% of their time for competence building, for example, courses, conferences, or learning new frameworks. Based on the documentation and the findings from *Phase* 1, we refined our interview guide. Our questions covered the following areas: *background, motivation, individual benefits/challenges, organizational benefits /challenges, virtual, hackathon organization, hackathon vs 20%.* Five semi-structured interviews were conducted by the first and the second author, with participants from across the organization, including one person organizing the hackathon, two that regularly attended them and two that had not participated in any hackathon. The interviews were conducted in Norwegian. Four of the interviews were recorded, and detailed notes were taken. In addition to data from the *Phase 2* interviews, we reviewed the qualitative data from *Phase 1* regarding outlining the effects of hackathons on productivity.

## 4. Results

In this section, we first report how a hackathon was conducted in the two companies, then we give an overview of the effects of the hackathon on productivity. Finally, we present results of hacking from home.

### 4.1. MobComp

**Who:** There is a strong culture for participation, and attendance rates are about 100%.
**How often:** Once a year.
**How long:** One dedicated week.
**Pre hackathon:** Ideas are submitted and put up for voting based on managerial input. The employees then vote for the three ideas they would like to join. Teams are formed based on the votes and managerial input to ensure suitable team sizes and diversity.
**Hackathon:** The hackathon lasts two days, and the teams decide how to organize the work. At the end of day two, there is a closing ceremony where all teams present their work. Previously, a winner was announced, but this was changed in the last hackathon.

**Table 1 Overview of hackathon effects on productivity**

| Level/ Dimension | Satisfaction & well-being: How fulfilled, happy and healthy people are | Performance: The outcome of a process | Activity: The count of actions or outputs | Communication & collaboration: How people talk and work together | Efficiency & flow: Doing work with minimal delays and interruptions |
|---|---|---|---|---|---|
| Individual | Fun from being with colleagues<br>Break from ordinary work<br>Building of new things<br>Doing work that one would never have time to do<br>Seeing what others can build<br>Feedback from peers | **MobComp**: Test new concepts quickly | Build critical knowledge fast<br>Develop many features | Get to know new people<br>Understand others' competences and fields of expertise (knowing who knows what)<br>A way for newcomers to meet people | Uninterrupted work |
| Organization | Build a sense of belonging to the company<br>Reduce stress - break from everyday work<br>Increased marketability and attraction of talent | **MobComp**: Ability to take advantage of emerging market opportunities<br>**FinCompTech**: Improve infrastructure and tools | **MobComp**: Test many product ideas fast | Knowing who knows who (build employee network)<br>Diversity<br>Teambuilding<br>**MobComp**: Speed up onboarding | A low number of handoffs during the hackathon |

The reason was to make it less of a competition and instead focus on celebrating all the ideas.

### 4.2. FinCompTech

**Who:** Everyone are invited (including consultants and marketing), attendance around 20%.

**How often:** Three times a year. Because of the pandemic, only once in 2020 (December).

**How long:** One dedicated week.

**Pre-hackathon:** Call for ideas/participation on Monday where anyone can pitch an idea and call for participants to their idea. Team formation starts and happens organically.

**Hackathon:** On Thursday, every idea with a minimum of two people moves forward. The teams self-organize into how, when, and where to work. At the end of Friday, there is a closing ceremony where hackathon teams present the results of the hackathon for the entire company. The ceremony involves an open vote, where a winner is crowned in four different categories: *the best user experience, the most technically impressive, the best business idea,* and *the best sustainable idea* (the latter added in 2020). The winner of each category receives a diploma and candy. Each participant can allocate the hours spent during the hackathon to their current project or a dedicated hackathon project.

### 4.3. Hackathon effects

To understand how hackathons affect productivity, we mapped our findings to the SPACE framework [14], and found effects for all five dimensions. Furthermore, we identified the effects on two distinct levels, i.e., *organizational* and *individual*. Our results are summarized in Table 1, where we also indicate which effects are only found in one of the companies. Next, we present our findings, according to the five dimensions.

**Satisfaction & well-being**. On the individual level, we find that hackathons enable developers to *Have fun with colleagues* and *build new things*. These two benefits are important for the developers *Satisfaction & well-being*, as one developer from MobComp mentions: "*I get positive feelings after the ending ceremony and presentation*". Having a break from everyday work was highly valued by individuals. One interviewee explained: *"We had a tough migration job just before the hackathon.... It was positive with a break after that period. And you know that the time is set aside for the hackathon, so you can just focus on that."*

Likewise, we examine the organizational effects. We see that hackathons strengthen the sense of belonging to the company: "*It gives the company a*

*possibility to build itself, and you can get smart people in."* Hackathons also provide a perk for recruitment: *"…webcam recording the football table so that you would just animate when you score a goal, and that was basically for fun. Eventually, we would use that in the recruiting processes to show students and convince them to come to the company."*

**Performance.** Informants in both companies mentioned the ability to try out various concepts: *"We were trying to explore a possible feature for the app"*, *"fun to test new products and get feedback"*. Recognizing the effects provided the companies with the ability to take advantage of emerging market opportunities through the innovation of new products, and both companies report that ideas gained through hackathons are being commercialized. In addition, hackathons increased quality because the code of existing software products was improved. As many as seven out of ten ideas were targeted towards infrastructure and support tooling in the most recent hackathon in MobComp.

**Activity.** We found that hackathons gave the opportunity to develop many features quickly. They also extended the ability to explore and build critical knowledge required for the development of the hackathon idea. The company, on the organizational level, receives many tested products and concepts.

**Communication & collaboration.** Informants from both companies gave insight into the importance of meeting new people: *"being able to work with someone you don't work with every day," "work with people across different teams,"* which then supported building internal networks. Working with others also gives the participants insight into the field expertise of other people and their competencies. As one developer explained: *"the good thing is that you do not only get insight into what other people do, but you also gain additional respect, 'really, that data is so hard to get by'"*. On the organizational level, we see that hackathons support the building of teams: *"In hackathons, what normally happens is that the main outcome, at least in our experience, tends to be the social aspect of it, the team building."* The growing personal network, the team building, was found to have a positive effect on the onboarding process. Informants from both companies mention that *"as a newcomer, it would be nice because you would instantly introduce yourself and you would get into that working environment with other people."*

**Efficiency & flow.** On the individual level, we found that the autonomy and self-management nature of the hackathons leads to higher *Efficiency & flow*. As explained by an informant in MobComp, *"One of the greatest advantages of hackathons is that you can jump over certain interactions that require quite a lot of time. Like discussions between a UX-designer and a developer or involving product owners"*. On the organizational level, this effect was noticeable in the low number of handoffs between teams and individuals during a hackathon.

## 4.4. Competence and learning

When analyzing our data, a new area emerged that influences productivity: *competence and learning*. This dimension is not part of SPACE. While one can argue that competence and learning is related to all the SPACE dimensions, we found it as a key benefit of hackathons and therefore describe it in Table 2.

**Table 2 Competence & learning**

| Level\Dimension | Competence and learning |
|---|---|
| Individual | Learning new skills (broader knowledge) |
| | Acquiring competence in new technology |
| | Honing existing skills (deeper knowledge) |
| | Increased understanding of how a feature can be developed |
| Organizational | Employees with broader knowledge |
| | Employees with deeper knowledge |

The increased competence into how features are developed was found to be valuable, especially for non-technical personnel; as mentioned by an advertisement operations manager at MobComp, *"You learn how much people can do in a very short period of time. And you learn that what you do in two days is rarely something different from what you would do in six months or one year."*

Working on a project not related to everyday work was seen as important, as explained by one developer at FinCompTech: *"If you choose an idea, somewhat outside of what you're working on every day, you might 'risk' learning something."* Another example is where an informant at MobComp used a hackathon to explore a new technology: *"…in one of the hackathons I set out to learn JavaScript and D3 for animations on webpages"*. He further explained how he used the hackathon to build knowledge before changing role: *"In one hackathon before I changed teams I worked on a mobile-application hack to learn about mobile development"*.

Informants also reported using the hackathon for deeper exploration of a particular technology. One informant from MobComp explained: *"There are two approaches for me in a hackathon, I either work on*

*something I am familiar with, or I focus on broadening my horizons.*" The individual learning results in a benefit on the organizational level, as a senior developer expressed: "*Yeah, I think ... it's a huge benefit for the company because people actually gain some new knowledge.*" Thus, the company is left with deeper and broader skilled personnel after the hackathon.

### 4.5. Hackathon challenges

Even though there are clear benefits of hackathons related to increased productivity, informants also mentioned challenges (Table 3). The challenge most frequently mentioned was that hackathons can add stress. Even though hackathons are mostly a welcomed break, the hackathon can be perceived as stressful. As a developer explained, *"you have pressure during your workday, and you need to get something finished and you look at hackathon as something in the way that is just disturbing you from important work."*. In FinCompTech, where hackathons are voluntary; organizers mention that stressed teams might be deterred from joining. This challenge is echoed in two challenges on the organizational side—stopped production for a time and increased employee stress.

Mismatch in skills and not having shared goals were mentioned by many. As many teams did not have knowledge redundancy (e.g., only one backend developer), being new in the company and not able to solve tasks as expected was stressful. Further, one interviewee explained how missing shared goals caused frustration, "*It was challenging that we had different goals we wanted to achieve. I wanted my product to be launched while the others wanted to expand their skills in machine learning.*"

**Table 3 Challenges of hackathons**

| Individual | Organization |
|---|---|
| Overhead of hackathon adds additional stress | Stops production for a time |
| **MobComp:** Mismatch in skills leads to slower development | **FinCompTech:** Employees do not partake in hackathon benefits |
| **MobComp:** Being unable to complete projects is somewhat frustrating | **MobComp:** Mismatch in skills leads to less time spent on emerging products |
| Lack of confidence is a barrier to joining | No-continuation plan leads to discontinued projects |
| Mismatch in individual goals for learning leads to disagreement in the teams | Employees are acquiring non-relevant skills |
| Poor communication and coordination due to mismatch in individual goals | Increased employee stress |

Although the employees were well aware that the hackathon was timeboxed, they still had a desire to finish their products. One participant explained how not finishing the work made him frustrated: "*I am not allowed to work on it. I have other work tasks. That's a downside with hackathons.*" This results in developers feeling dissatisfied as their goals remain unfulfilled, as one explained: "*You know, the worst feeling you can have is when you write code and after some time it gets forgotten and no one ever uses it.*"

### 4.6. Hackathons during the pandemic

Due to the Covid-19 pandemic, both companies organized virtual hackathons. We found that going virtual affected the hackathons. In MobComp, where everyone was expected to participate, we observed that idea generation, idea voting, and attendance were similar to the pre-Covid hackathons. However, at FinCompTech where there was no management involvement in picking ideas or forming teams, there was a noticeable reduction in the number of ideas and hackathon pitches. At FinCompTech this hackathon was the first in a year (the previous once had been cancelled because of the pandemic).

Despite the pre-hackathon phase change at MobComp, there was no significant impairment in developing or demoing the technical solution for any of the companies, as one from MobComp explained: *".. from a technical point of view, to just develop something, you can always use Slack, or whatever, or just Google meet and call."* However, we found a change in how tasks were solved. In the virtual hackathons, the employees started by dividing the work among themselves, then worked individually, coordinating through scheduled meetings and Slack, and used virtual meeting rooms for discussions and planning: (MobComp) *"We did the first meeting when we split the work. So, after a few hours, we met again. We have this approach; let's just give it a try. Then a few hours later, we had another meeting."* Dividing tasks early, and mainly working on one's own tasks cased problems when a team member could not finish his tasks. If one was delayed, the demo was affected. While the progress was tracked quite often in the virtual hackathon and they used Slack, the participants work more closely together in physical hackathons where they talked to each other and took breaks together. Further, in the pre pandemic sessions the whole team would stay in the office to socialize and support each other while eating pizza until everything was finished, team members would now log off and end the day if they had no more tasks to do.

A reduction in the quality of having fun with colleagues was also reported: "*It was not as much fun as it used to be. Hackathons are usually more fun physically, than over video during Covid, in my own home.*" One from MobComp commented: "*I did not join last year because it was virtual. I wanted to go back to the office and meet people and did not want to join another virtual work session.*" Most employees mentioned that they would prefer to have the hackathon in the same physical location to socialize and have fun with colleagues.

Hackathon as a break from ordinary work was seen as a positive by several, e.g. after stressful periods such *as* migrations. It also enabled meeting others than those you were working with daily, as one from MobComp said," Especially *in these pandemic times. You know we are working from home. So, we want to meet people … and hackathons are one way to just, you know, meet and collaborate.*"

Lastly, disagreements within a team due to a mismatch in individual goals seem to be harder to solve when they are held virtually. One reason is that it is harder to identify when participants within a team have different goals and speake less to each other. We did not find any teams that used time to discuss expectations or individual goals.

## 5. Discussion

We studied hackathons in two companies to understand how the practice relates to productivity and what changed when everyone was forced to work from home. At MobComp, the practice was yearly and perceived as mandatory, resulting in everyone participating. It was a key practice for building the company culture. Meanwhile, at FinCompTech, the hackathons were organized three times per year, and less than one-third of the employees participated. Moreover, in FinCompTech employees had 20% of their work hours allocated for personal development, learning, and working on projects of their choosing. Unlike findings at Atlassian [17], where the hackathon and 20% time were found to cover different aspects of innovation and learning, the 20% practice was seen by some to be redundant to hackathons at FinCompTech.

While FinCompTech had no restrictions on choosing ideas, MobComp had a strategy of selecting and managing topics and teams to increase diversity through cross-company teams. Our findings corroborate [2], which states that a hackathon can be a mechanism for connecting marginally connected parts of the company. In comparison, at FinCompTech where the teams created themselves, the teams had a much higher degree of homogeneity and sometimes consisted of existing teams.

While MobComp focused more on competence building and new features or products, FinCompTech focused more on building up or improving internal tooling, infrastructure, pipelines etc. for supporting the software development process. Making it more efficient to develop software could potentially improve the productivity of the company. One reason for a more internal focus at FinCompTech could be that the company was much larger, and then the internal tools would be needed by many.

### 5.1. Hackathons and productivity

We now discuss our first research question: "*How do hackathons affect productivity in large-scale agile?*" We relied on the SPACE framework [14] to describe the benefits or the effects of Hackathons. SPACE provided us with a way to systematically report factors affecting productivity and helped illuminate tradeoffs between effects.

Consonant with [17] and [8], we found that providing dedicated days for developers may lead to innovation and new product development. Further, we discovered that hackathons improve satisfaction and well-being, performance, activity, communication and collaboration, and efficiency and flow. These effects were found on the individual and organizational level (Table 1). Therefore we argue that the SPACE framework [14] is suitable for explaining how hackathon affects productivity.

The benefits are related. Working together with new people improves a person's network, supports onboarding of new people and is important for having fun. Working on internal tooling gives a break from everyday work, it improves the development process; employees become more satisfied and learn something new. However, we also found tradeoffs between some of the benefits reported. For example, relating to building new things vs teambuilding, our findings show that tailoring teams for diversification might reduce the team's performance, corroborating Falk Olesen and Halskov [1], where the skills wanted by individuals did not match those of the organization.

Challenges found in the two cases impacted the benefits. Some employees were frustrated when their product was not continued, this confirms a challenge with hackathon continuation as reported in previous research [5, 9]. In MobComp, this challenge can be attributed to the fact that the hackathons focused on developing new and innovative products. In contrast employees at FinCompTech worked more on internal tooling, which was easier to utilize and follow up on than new products with uncertain revenue potential.

Furthermore, employees at FinCompTech had 20% time and could continue the work.

We also found additional challenges, such as taking time away from everyday work, whether it was good timing or not, corroborated by [8]. In addition, personal desires and goals could also pose a challenge where teams could not agree on common goals. Finally, we found that while new business opportunities might be important for the company, having fun and being with colleagues were more critical for the individual. This might indicate that steering what ideas were allowed to be pitched might reduce the effect of a hackathon

We found that learning and networking are important benefits, as corroborated by [1, 23]. The SPACE framework does not directly highlight *competence & learning* as we have found essential for developers. We also recognize the need in large-scale agile companies for extra-curricular activities in addition to the regular agile activities, this corroborates SAFe's [26] mention of the need to allocate time for working on activities that will not be prioritized continuously. We thus argue that large-scale agile should include hackathons and other practices to allow for a change of scenery and learning new competencies required to develop and utilize new technology, as well as breaking down communication barriers between autonomous agile teams.

### 5.2. Hacking from home

When answering our second research question: "*How do developers experience hackathons held virtually when working from home?*", our results suggest that the shift from physical to virtual hackathons in large-scale agile does not impact the ability to develop the technical solution or the consequent demo of the solution. However, we found that the number of hours spent on the hackathon decreased compared to co-located hackathons, as people logged off when their part of the work was done, and not hanging around as they would have done. The tasks were also more divided between the participants. Dividing the tasks among team members reduced the need for collaboration, which makes coordination easier but might weaken the team [27]. However, the employees claimed to produce high-quality demo prototypes and were happy with their outcomes. Furthermore, while people enjoyed working with new people, we found that hackathons are less fun when conducted virtually since physical hackathons provide more encouragement and room for socializing. As one explained, "*it is less fun*". We also found that virtual hackathons reduced attendance if that attendance was voluntarily. These results show that virtual hackathons do not provide the same level of benefits compared to co-located hackathons and therefore requires extra effort.

### 5.3. Limitations

A first limitation is that we used a multiple case design. Therefore, the general criticisms of case studies, such as uniqueness and special access to key informants, may also apply in our study. However, the rationale for choosing the two companies was that they organized hackathons in two different ways: in one, everyone participated, and it was held annually, on the other around 20% participated, and it was organized three times a year. As many phenomena were reported in both companies, it is likely that other large-scale companies report the same phenomena and that the conclusions in this study will prove useful. Another possible limitation is that much of the data collection and analysis was based on semi-structured interviews. The consequence of this limitation is that the results have been influenced by our interpretation of the phenomena observed and investigated.

## 6. Conclusion and future work

Our results suggest that virtual hackathons in large-scale agile organizations improve productivity on the individual and organizational level. First, hackathons improve the satisfaction and well-being of developers, as it is a fun practice and provides a break from everyday work. Second, it strengthens the company culture and improves performance as many ideas are tested. Third, the process increases activity as ideas are developed and demonstrated quickly. Fourth, it increases communication and collaboration as the social network is strengthened and people are on-boarded more quickly. Fifth, it increases efficiency and flow as people learn how to complete work or make progress with minimal interruption or delay. Finally, a hackathon is a resource for learning and building broader and deeper competence.

Even though hackathons provide the organization with new products, refreshed employees and a more competent workforce, practitioners should be aware that challenges include additional stress and dissonance in hackathon teams due to differences in experience and personal goals.

As companies such as Twitter, Spotify, Facebook and Salesforce have developed work-from-anywhere strategies, it is likely that hackathons in the future will be conducted in a virtual setting. Therefore, future research should focus on collecting more data about virtual hackathons in other large-scale agile organizations to develop insights into how virtual hackathons can be organized and improved. More specifically, there is a need to research how teams work together as we have found that team members collaborated less in virtual hackathons and have less fun. There is also a need to develop metrics as suggested in the SPACE framework.

## 7. Acknowledgements

This research is funded by the Research Council of Norway through the 10xTeams project (grant 309344).

## 8. References

[1] J. Falk Olesen, K. Halskov, 10 years of research with and on hackathons, in: Proceedings of the 2020 ACM designing interactive systems conference, 2020, 1073-1088.
[2] A. Nolte, E.P.P. Pe-Than, A. Filippova, C. Bird, S. Scallen, J.D. Herbsleb, You Hacked and Now What? - Exploring Outcomes of a Corporate Hackathon, Proceedings of the ACM on Human-Computer Interaction, 2 (2018) 1-23.
[3] G. Briscoe, C. Mulligan, Digital innovation: The hackathon phenomenon, (2014).
[4] G. Valença, N. Lacerda, C.R.B. de Souza, K. Gama, A Systematic Mapping Study on the Organisation of Corporate Hackathons, in: 2020 46th Euromicro Conference on Software Engineering and Advanced Applications (SEAA), IEEE, 2020, 421-428.
[5] M. Komssi, D. Pichlis, M. Raatikainen, K. Kindström, J. Järvinen, What are hackathons for?, IEEE Software, 32 (2014) 60-67.
[6] A. Sablis, D. Smite, N. Moe, Team-external coordination in large-scale software development projects, Journal of Software: Evolution and Process, 33 (2021) e2297.
[7] M. Mikalsen, N.B. Moe, V. Stray, H. Nyrud, Agile digital transformation: a case study of interdependencies, (2018).
[8] H.T. Barney, N.B. Moe, T. Dybå, A. Aurum, M. Winata, Balancing individual and collaborative work in agile teams, in: International Conference on Agile Processes and Extreme Programming in Software Engineering, Springer, 2009, 53-62.
[9] A. Nolte, I.-A. Chounta, J.D. Herbsleb, What Happens to All These Hackathon Projects? Identifying Factors to Promote Hackathon Project Continuation, Proceedings of the ACM on Human-Computer Interaction, 4 (2020) 1-26.
[10] M. Flores, M. Golob, D. Maklin, M. Herrera, C. Tucci, A. Al-Ashaab, L. Williams, A. Encinas, V. Martinez, M. Zaki, How can hackathons accelerate corporate innovation?, in: IFIP International Conference on Advances in Production Management Systems, Springer, 2018, 167-175.
[11] C. Kollwitz, B. Dinter, What the hack?–towards a taxonomy of hackathons, in: International Conference on Business Process Management, Springer, 2019, 354-369.
[12] D. Graziotin, F. Fagerholm, Happiness and the Productivity of Software Engineers, in: C. Sadowski, T. Zimmermann (Eds.) Rethinking Productivity in Software Engineering, Apress, Berkeley, CA, 2019, 109-124.
[13] S. Nerur, V. Balijepally, Theoretical reflections on agile development methodologies - The traditional goal of optimization and control is making way for learning and innovation, Communications of the Acm, 50 (2007) 79-83.
[14] N. Forsgren, M.-A. Storey, C. Maddila, T. Zimmermann, B. Houck, J. Butler, The SPACE of Developer Productivity: There's more to it than you think, Queue, 19 (2021) 20-48.
[15] R. Ulfsnes, V. Stray, N.B. Moe, D. Šmite, Innovation in Large-scale agile--Benefits and Challenges of Hackathons when Hacking from Home, in: XP2021 Workshop, 2021.
[16] C.W. Langfred, The paradox of self-management: Individual and group autonomy in work groups, Journal of Organizational Behavior, 21 (2000) 563-585.
[17] N.B. Moe, S. Barney, A. Aurum, M. Khurum, C. Wohlin, H.T. Barney, T. Gorschek, M. Winata, Fostering and sustaining innovation in a fast growing agile company, in: Proceedings of the 13th international conference on Product-Focused Software Process Improvement, Springer-Verlag, Madrid, Spain, 2012, 160-174.
[18] M.Y. Lee, A.C. Edmondson, Self-managing organizations: Exploring the limits of less-hierarchical organizing, Research in organizational behavior, 37 (2017) 35-58.
[19] D. Smite, N.B. Moe, G. Levinta, M. Floryan, Spotify guilds: how to succeed with knowledge sharing in large-scale agile organizations, IEEE Software, 36 (2019) 51-57.
[20] K. Dikert, M. Paasivaara, C. Lassenius, Challenges and success factors for large-scale agile transformations: A systematic literature review, Journal of Systems and Software, 119 (2016) 87-108. http://dx.doi.org/10.1016/j.jss.2016.06.013
[21] T. Sporsem, A. Tkalich, N.B. Moe, M. Mikalsen, Understanding Barriers to Internal Startups in Large Organizations: Evidence from a Globally Distributed Company, in: ICGSE, 2021.
[22] C. Jaspan, C. Sadowski, No single metric captures productivity, in: Rethinking Productivity in Software Engineering, Springer, 2019, 13-20.
[23] L. Paganini, K. Gama, A preliminary study about the low engagement of female participation in hackathons, in: Proceedings of the IEEE/ACM 42nd International Conference on Software Engineering Workshops, 2020, 193-194.
[24] R.K. Yin, Case study research and applications: Design and methods, Sage publications, 2017.
[25] J. Saldaña, The coding manual for qualitative researchers, Sage, 2015.
[26] D. Leffingwell, SAFe 4.5 Reference Guide: Scaled Agile Framework for Lean Enterprises, Addison-Wesley Professional, 2018.
[27] N.B. Moe, T. Dingsøyr, T. Dybå, A teamwork model for understanding an agile team: A case study of a Scrum project, Information and Software Technology, 52 (2010) 480-491.